\documentclass{raa}

\usepackage{graphicx,times}
\usepackage{natbib}
\usepackage{amssymb,amsmath}
\bibpunct{(}{)}{;}{a}{}{,}

\usepackage[pagebackref=true]{hyperref}

\usepackage{epsfig,amsmath,natbib}
\usepackage{color}
\usepackage{xcolor}

\usepackage{pifont}

\usepackage{makecell}
\usepackage{array}
\usepackage{booktabs}
\usepackage{bm}
\usepackage{multirow}

\newcommand{\Mpc}{\,{\rm Mpc}}

\begin{document}

   \title{Reionizing islands with inhomogeneous recombinations}


   \author{Runyu Zhu\inst{1,2}, Yidong Xu\inst{1}$^*$, Bin Yue\inst{1}, Xuelei Chen\inst{1,2,3,4}$^*$
   \footnotetext{$*$ Corresponding author}
   }

   \institute{ 
National Astronomical Observatories, Chinese Academy of Sciences, 20A Datun Road, Beijing 100101, P. R. China; {\it xuyd@nao.cas.cn}; {\it xuelei@cosmology.bao.ac.cn}\\
\and
School of Astronomy and Space Science, University of Chinese Academy of Sciences, Beijing 100049, P. R. China
\\
\and
Department of Physics, College of Sciences, Northeastern University, Shenyang, 110819, China\\
\and 
Center of High Energy Physics, Peking University, Beijing 100871, China\\
\vs \no
   {\small Received 20XX Month Day; accepted 20XX Month Day}
}

\abstract{Observations are beginning to constrain the history of the epoch of reionization (EoR). Modeling the reionization process is indispensable to interpret the observations, to infer the properties of ionizing sources, and to probe the various astrophysical processes from the observational data. Here we present an improved version of the semi-numerical simulation {\tt islandFAST}, by incorporating inhomogeneous recombinations and a corresponding inhomogeneous ionizing background, and simulate the reionization process of neutral islands during the late EoR. We find that the islands are more  
fragmented in models with inhomogeneous recombinations than the case with a homogeneous recombination number. 
In order to investigate the effects of basic assumptions in the reionization modeling, we compare the results from {\tt islandFAST} with those from {\tt 21cmFAST} for the same assumptions on the ionizing photon sources and sinks, to find how the morphology of the ionization field and the reionization history
depend on the different treatments of these two models.
Such systematic bias should be noted when interpreting the upcoming observations. 
\keywords{(cosmology:) dark ages, reionization, first stars -- (galaxies:) intergalactic medium -- (cosmology:) large-scale structure of Universe -- galaxies: high-redshift
}
}

   \authorrunning{R. Zhu et al. }            
   \titlerunning{islandFAST with inhomogeneous recombinations}  
   \maketitle

%
\section{Introduction}           
\label{sect:intro}

The cosmic hydrogen reionization process is a  complex interplay between the early galaxy formation and evolution of the intergalactic medium (IGM) \citep{Ciardi2005review}.
While there are few direct observations of the reionizing IGM, large amount of efforts have been made to constrain the reionization history with indirect observations.
For example, the measurement of Thomson optical depth of the cosmic microwave background (CMB) from Planck data suggests a mid-point reionization redshift of $z_{\rm{re}} = 7.7\pm0.7$ \citep{Planck2020}. 
Observations of the ``Gunn–Peterson'' trough
on spectra of high-redshift quasars (e.g. \citealt{Fan2006AJ, Bouwens2015ApJ}), and the fraction of dark pixels in the Ly$\alpha$ and Ly$\beta$ forests (\citealt{McGreer2015MNRAS}), indicate that reionization completes approximately at $z\sim6$. The Ly$\alpha$ damping wing (e.g. \citealt{Banados2018Natur}), the Ly$\beta$ forest in quasar spectra \citep{Zhu2022ApJ}, the Lyman-break galaxies (e.g. \citealt{Mesinger2015MNRAS}), the clustering of Ly$\alpha$ emitters \citep{Sobacchi2015MNRAS}, Ly$\alpha$ luminosity function (e.g. \citealt{Morales2021ApJ}), as well as Ly$\alpha$ equivalent-width distributions \citep{Jung2020ApJ}, have all been utilized to constrain the neutral fraction at the late epoch of reionization (EoR). Recently, it has also been proposed that measurements of the dispersion measure of fast radio bursts will be a potentially promising probe to the reionization history (e.g. \citealt{Fialkov2016JCAP, Dai2021JCAP, Hashimoto2021MNRAS}).

More direct constraints will come from observations of high-redshift galaxies by the James Webb Space Telescope (JWST) (e.g.  \citealt{Wang2022arXiv221204476W, Robertson2022arXiv221204480R, Curtis-Lake2022arXiv221204568C}) 
that directly probe the ionizing sources, and the 21 cm observations of neutral hydrogen that directly probe the ionization and heating process of the IGM \citep{Furlanetto2006PhR,Pritchard2012RPPh}.
Detection of the global 21 cm spectrum is still unsettled, with conflicting results from the Experiment to Detect the Global EoR Signature (EDGES, \citealt{Bowman2018Natur}) and 
the Shaped Antenna measurement of the background RAdio Spectrum (SARAS) experiment \citep{Singh2022NatAs}. 
Improved global spectrum experiments will put some constraints on the first galaxies and the level of a possible early radio background during the cosmic dawn \citep{Bevins2022}, while the recent upper limit on the 21 cm power spectrum from the Hydrogen Epoch of Reionization Array (HERA; \citealt{HERAlimit2022}) has put a lower bound on early X-ray heating level during the EoR \citep{HERAconstrain2022}. 
Combination of these two probes have put the tightest constraints on the earliest galaxies to date \citep{Bevins2023}.

To interpret the observations, a good theoretical model is needed, and the uncertainty in the model needs to be well understood \citep{HERAconstrain2022}.
One widely-accepted model for reionization is the so-called ``bubble model'' \citep{Furlanetto2004ApJ}, which assumes a bubble-like topology for ionized regions. The ionization state of a given point is determined by 
comparing the
 required number of ionizing photons with the number of such photons produced in the surrounding region. Using the excursion set model of density field fluctuations \citep{Bond1991ApJ, Lacey1993MNRAS}, the size distribution of ionized bubbles is computed by solving for the first up-crossing distribution with respect to a density barrier for ionization. This algorithm becomes the basis of many semi-numerical simulations (e.g. \citealt{Mesinger2007ApJ, Zahn2007ApJ, Geil2008MNRAS, Alvarez2009ApJ, Mesinger2011MNRAS, Zhou2013RAA, Lin2016MNRAS, Paranjape2016MNRAS}). 
After percolation of ionized bubbles, the topology of the ionization field is no longer bubble-like. Then
the island model was developed to improve the model performance by assuming an island-like topology for the remaining neutral regions during the late EoR \citep{Xu2014ApJ}.
A semi-numerical simulation {\tt islandFAST} was developed based on the island model \citep{Xu2017ApJ}.

The original {\tt islandFAST} adopted a homogeneous model for the evolving ionizing background and a constant recombination rate.
However, the recombination rate depends on the density which have spatial variations, 
and there are large-scale spatial fluctuations in the ionizing background \citep{Davies2016MN,Becker2018}.
In the present work, we improve the {\tt islandFAST} by incorporating an inhomogeneous model for the recombination rate, and model the ionizing background which arises from the inhomogeneous distribution of both sources and sinks of the ionizing photons. 
We investigate the effect of inhomogeneous recombinations and ionizing background on the ionization history and the morphology of the ionization field. We also compare {\tt islandFAST} with {\tt 21cmFAST} \citep{Park2019MNRAS} for the late EoR, assuming the same astrophysical parameters, in order to investigate 
the uncertainty of the ionization fields and 
reionization history in current semi-numerical models.

This paper is organized as follows. In Section \ref{sect:islandFAST} we present the improved model and algorithm of the {\tt islandFAST} program with inhomogeneous recombinations and ionizing background. Section \ref{sect:nrec} illustrates the effects of inhomogeneous recombinations on the ionization history and morphology of the ionization field. The differences between {\tt islandFAST} and {\tt 21cmFAST} models are compared and analyzed in Section \ref{sect:model}. We summarize our main results and implications in Section \ref{sect:sum}.
Throughout our analysis, we adopt the $\Lambda$CDM cosmology based on Planck 2018 results \citep{Planck2020}: $\Omega_{\rm{\Lambda}} = 0.685$, $\Omega_{\rm m} = 0.315$, $\Omega_{\rm b} = 0.048$, $H_{\rm 0} = 67.4\, \rm {km\, s^{-1} Mpc^{-1}}$, $\sigma_{\rm 8} = 0.811$, and $n_{\rm s} = 0.965$.

\section{{\tt islandFAST} with inhomogeneous recombinations}
\label{sect:islandFAST}

\subsection{island model with a homogeneous ionizing background}
\label{sect:old_version}

Here we first briefly review the basic algorithm of the island model and the original {\tt islandFAST} with a homogeneous ionizing background, but refer interested readers to \citet{Xu2014ApJ} and \citet{Xu2017ApJ} for more details. 

During the early stages of reionization, one can neglect the contribution from the ionizing background, and the reionization process can be well described by the bubble model \citep{Furlanetto2004ApJ}. 
Assuming the production of ionizing photons to be proportional to the collapse fraction, the ionization criteria, referred to as the ``bubble barrier" in excursion set theory, can be written as:
\begin{equation}\label{Eq.bubblebarrier}
\begin{split}
\zeta f^{R}_{\mathrm{coll}}\left(\delta_{M}; M_{\rm min}, z\right) \geq 1+\Bar{n}_{\rm rec},
\end{split}
\end{equation}
where $f^{R}_{\mathrm{coll}}(\delta_{M}; M_{\rm min}, z)$ is the collapse fraction of a region with mass scale $M$ and mean overdensity $\delta_{M}$ at redshift $z$, $R$ is the smoothing scale corresponding to $M$, $M_{\rm min}$ is 
the minimum mass of halos that contribute to ionizing photons,
$\Bar{n}_{\mathrm{rec}}$ is the mean recombination number for each ionized hydrogen atom, and $\zeta\equiv f_{\star}N_{\gamma/\rm H}f_{\rm{esc}}$ is the ionizing efficiency parameter, in which $f_{\star}$, $N_{\gamma/\rm H}$, and $f_{\mathrm{esc}}$ are the star formation efficiency, the number of ionizing photons emitted per H atom in stars, and the escape fraction of ionizing photons, respectively.
Based on the framework of excursion set theory (e.g. \citealt{Bond1991ApJ, Lacey1993MNRAS}, the ionized regions are identified by the \textit{first} up-crossings of the bubble barrier, as in {\tt 21cmFAST} \citep{Mesinger2011MNRAS}.  

After percolation of the ionized bubbles,  a global ionizing background is gradually set up, and its contribution should be accounted. The {\tt islandFAST} adopts an inverse topology at the late stages as the neutral islands are more isolated after bubble percolation. It first identifies host islands with a neutral criterion including an ionizing background, and then takes into account the bubbles-in-island effect, i.e. bubbles formed in large islands, by applying an ionization criterion within the host islands. The condition for a region of mass scale $M$ to keep from fully ionized at redshift $z$, or the ``island barrier'', is
\citep{Xu2014ApJ}:
\begin{equation}\label{Eq.islandbarrier}
\begin{split}
     \zeta f^{R}_{\mathrm{coll}}\left(\delta_{M} ; M_{\rm min}, z\right) + \frac{\Omega_m}{\Omega_b} \frac{N_{\mathrm{back}} m_{\mathrm{H}}}{M X_{\mathrm{H}}} < 1+\Bar{n}_{\mathrm{rec}},
\end{split}
\end{equation}
where $N_{\mathrm{back}}$ is the number of consumed background ionizing photons by the region under consideration, $m_{\mathrm H}$ is the mass of a hydrogen atom, and $X_{\mathrm H}$ is the 
mass fraction of hydrogen.
The ionizing background carves an island outside in, and
the number of background ionizing photons consumed by the island is approximately calculated by
\begin{equation}
\begin{split}
N_{\mathrm{back}}\approx\frac{4 \pi}{3}\left(R_{\rm Ii}^3-R_{\rm If}^3\right) \bar{n}_{\mathrm{H}}\left(1+\Bar{n}_{\mathrm{rec}}\right),
\end{split}
\end{equation}
where $R_{\rm Ii}$ and $R_{\rm If}$ are the initial and final comoving scale of the island, at the redshift interval under consideration, respectively, and $\bar{n}_{\mathrm {H}}$ is the average number density of hydrogen in the IGM. 
The scale change of the island is calculated by integrating the shrinking process up to the ``background onset redshift''
$z_{\mathrm {back}}$, i.e.
\begin{equation}\label{Eq.DeltaR_homo}
\begin{split}
   \Delta R_{\rm I} \equiv R_{\rm Ii}-R_{\rm If}=\int_{z}^{z_{\text {back}}} \frac{F(z)}{\bar{n}_{\mathrm{H}}\left(1+\bar{n}_{\mathrm{rec}}\right)} \frac{\mathrm{d} z}{H(z)(1+z)^{3}}.
\end{split}
\end{equation}
Here $F(z) = n_{\mathrm \gamma}(1+z)^{3}c/4$ is the physical photon number flux of the ionizing background, in which $n_{\gamma}(z)$ is the comoving number density of background ionizing photons. 
Assuming a homogeneous ionizing background, we have
\begin{equation}\label{Eq.ngamma_homo}
\begin{split}
    n_{\gamma}(z)=\int_{z} \bar{n}_{\mathrm{H}}\left|\frac{\mathrm{d}f^\infty_{\text {coll}}\left(z^{\prime}\right)}{\mathrm{d} z^{\prime}}\right|
    \zeta\,\exp\left[-\frac{l\left(z, z^{\prime}\right)}{\lambda_{\operatorname{mfp}}(z)}\right]\left(1-f_{\mathrm {HI}}^{\mathrm {host}}\right)\mathrm{d} z^{\prime},
\end{split}
\end{equation}
where $l\left(z,z^{\prime}\right)$ is the physical distance between the source at redshift $z'$ and the redshift $z$ under consideration, and $\lambda_{\mathrm {mfp}}$ is the physical mean free path (MFP) of the background ionizing photons. Here $f_{\rm HI}^{\rm host}$ is the volume fraction occupied by host islands, and we add the factor $\left(1-f_{\mathrm {HI}}^{\mathrm {host}}\right)$ because the background ionizing photons are contributed by sources in ionized regions.

The MFP of ionizing photons is limited by two kinds of absorbers, i.e. the large-scale under-dense neutral islands, and the small-scale over-dense absorbers that are not resolved in the semi-numerical simulation. The effective MFP can be written as
\begin{equation}\label{LAMBDA_MFP}
\begin{split}
    \lambda_{\mathrm {mfp}}^{-1}=\lambda_{\mathrm {HI}}^{-1}+\lambda_{\mathrm {abs}}^{-1},
\end{split}
\end{equation}
where $\lambda_{\mathrm {HI}}$ and $\lambda_{\mathrm {abs}}$ are the MFP limited by neutral islands and by small-scale absorbers, respectively.
$\lambda_{\mathrm {HI}}$ is self-consistently computed from the ionization field using the mean-free-path algorithm, while an empirical fitting formula for an evolving $\lambda_{\mathrm {abs}}$ \citep{Songaila2010ApJ} is used in the original {\tt islandFAST}.

During the island stage of reionization \citep{Chen2019ApJ}, {\tt islandFAST} adopts a two-step filtering algorithm. It first finds host islands by identifying regions {\it first} down-crossing the island barrier, and then finds bubbles in islands by applying again the bubble barrier within host islands. The intensity of the ionizing background is solved simultaneously with the ionization field.

\subsection{{\tt islandFAST} with inhomogeneous recombinations}

Considering the large-scale clustering of both sources and sinks, and the unresolved small-scale density fluctuations, the ionizing background, recombination rate, and the reionization process are all inhomogeneous in nature. Here we improve the {\tt islandFAST} by including inhomogeneous recombinations as well as a spatially-varying ionizing background.
The physical properties of small-scale ionizing photon sinks, and their contribution to the IGM opacity and effects on the ionizing background, were studied using radiative hydrodynamic simulations \citep{McQuinn2011,Rahmati2018,Nasir2021}.
Effects of inhomogeneous recombinations on the large-scale reionization can be incorporated by using a subgrid model for small-scale density distribution \citep{Sobacchi2014MNRAS}, or applying a clumping factor-overdensity correlation fitted to a high-resolution N-body simulation \citep{MaoClumping2020}.
To ease model comparison with {\tt 21cmFAST}, here we follow
the basic formalisms for recombinations in \citet{Sobacchi2014MNRAS}. 
The basic framework of {\tt islandFAST}, with the bubble barrier and island barrier, still applies, but $n_{\rm rec}(\boldsymbol{x}, z)$, $\lambda_{\rm mfp}(\boldsymbol{x}, z)$, $n_\gamma(\boldsymbol{x}, z)$, and $\Delta R_{\rm I}(\boldsymbol{x}, z)$ are now allowed to vary with position $\boldsymbol{x}$, as detailed below.

\subsubsection{Inhomogeneous recombinations and MFP of ionizing photons}

Small-scale absorbers play a significant role in modulating the reionization process (e.g. \citealt{McQuinn2007MNRAS, Alvarez2012, Sobacchi2014MNRAS}),
and the effects are more important during the late EoR when typical ionized regions are larger than the MFP of ionizing photons \citep{Wu2022ApJ}.
The growth of large HII regions would be impeded by dense structures with densities close to the self-shielding threshold \citep{Miralda2000ApJ,Furlanetto2005MNRAS}.
Unfortunately, it is challenging to resolve the distribution of small-scale absorbers in large-scale reionization simulations. 
An empirical formula for the gas density distribution based on numerical simulations has been developed by \citet{Miralda2000ApJ}, which gives the volume fraction of gas at overdensity $\Delta \equiv n_{\mathrm {b}}/\bar{n}_{\mathrm {b}}$ (hereafter MHR00 distribution):
\begin{equation}
\begin{split}
    P_{\mathrm v}(\Delta,z)=A\,{\mathrm {exp}}\left[-\frac{\left(\Delta^{-2/3}-C_0\right)^{2}}{2\left(2\delta_0/3\right)^{2}}\right]\Delta^{-\beta}.
\end{split}
\end{equation}
Here we follow \citet{Sobacchi2014MNRAS} and adopt $\delta_{\mathrm 0} = 7.61/(1+z_{\mathrm {eff}})$ evaluated at an effective redshift $(1+z_{\mathrm {eff}}) \equiv (1+z)\Delta_{\mathrm {cell}}^{1/3}$, in which $\Delta_{\mathrm {cell}}$ is the mean overdensity for each simulation cell, and $\beta = 2.5$ for high redshifts. $A$ and $C_{0}$ are two constants at each redshift, determined by normalizing the total volume and mass fraction to unity. 
Then the local sub-grid recombination rate can be calculated by integrating over the entire density distribution \citep{Park2019MNRAS}:
\begin{equation}\label{Eq.DNREC_DT}
\begin{split}
    \frac{{\mathrm d}n_{\mathrm {rec}}}{{\mathrm d}t}(\boldsymbol{x},z)=\bar{n}_{\mathrm H}\alpha_{\mathrm B}\Delta_{\mathrm {cell}}^{-1}\int_{0}^{180}\left[1-x_{\mathrm {HI}}(\Delta)\right]^{2}P_{\mathrm v}(\Delta,z)\Delta^{2}{\mathrm d}\Delta.
\end{split}
\end{equation}
Here $\alpha_{\rm B}$ is the case B recombination coefficient, and
$x_{\mathrm {HI}}(\Delta)$ is the local neutral fraction associated with the sub-grid overdensity $\Delta$, which is calculated assuming local photoionization equilibrium \citep{Sobacchi2014MNRAS}.
The total recombination number per baryon, averaged over the smoothing scale $R$ under consideration, can be integrated over time-steps:
\begin{equation}\label{NREC_CALCULATION}
\begin{split}
    \bar{n}_{\mathrm {rec}}(\boldsymbol{x},z,R)=\left<\int_{z_{\mathrm {ion}}}^{z}\frac{{\mathrm d}n_{\mathrm {rec}}}{{\mathrm d}t}\frac{{\mathrm d}t}{{\mathrm d}z}{\mathrm d}z\right>_R,
\end{split}
\end{equation}
where $z_{\mathrm {ion}}$ is the redshift of first ionization of each ionized cell. The homogeneous recombination number in the bubble barrier (Eq.~\ref{Eq.bubblebarrier}) and island barrier (Eq.~\ref{Eq.islandbarrier}) is substituted with this inhomogeneous number.

Assuming that small-scale absorbers are dominated by self-shielding systems with a uniform shape, the MHR00 distribution of local overdensities also sets the MFP limited by these absorbers. 
The volume fraction of the IGM occupied by absorbers is $Q_{\mathrm {ss}}=\int_{\Delta_{\mathrm {ss}}}^{\infty}P_{\mathrm v}(\Delta,z){\mathrm d}\Delta$, where $\Delta_{\mathrm {ss}}$ is the self-shielding threshold of overdensity. Then the MFP from small absorbers is given by
\begin{equation}\label{Eq.LAMBDA_ABS}
\begin{split}
    \lambda_{\mathrm {abs}}=\lambda_0Q_{\mathrm {ss}}\left(\Delta_{\mathrm {ss}}\right)^{-2/3},
\end{split}
\end{equation}
where $\lambda_0$ is the normalization factor and  $\lambda_0 H=60\,\mathrm {km\ s}^{-1}$ is chosen to account for the cumulative opacity from lower-density systems while matching lower redshift observations \citep{Furlanetto2005MNRAS}. The self-shielding threshold depends on the photoionization rate of the ionizing background $\Gamma_{\rm HII}$ and the IGM temperature $T$, given by \citep{Sobacchi2014MNRAS}:
\begin{equation}\label{DSS}
\begin{split}
    \Delta_{\mathrm{ss}}=27 \times\left(\frac{T}{10^{4} \mathrm{~K}}\right)^{0.17}\left(\frac{1+z}{10}\right)^{-3}\left(\frac{\Gamma_{\mathrm{H} \mathrm{II}}}{10^{-12} \mathrm{~s}^{-1}}\right)^{2/3}.
\end{split}
\end{equation}
Here we adopt $T=10^{4}\, \mathrm{K}$ for the gas in ionized regions.
Eq.(\ref{Eq.LAMBDA_ABS}) provides a good approximation for the MFP due to unresolved absorbers at the late EoR \citep{Sobacchi2014MNRAS}.
Both $\Delta_{\rm {ss}}$ and $\Gamma_{\rm HII}$ are spatial-dependent, thus $\lambda_{\rm {abs}}$ and the total MFP $\lambda_{\rm mfp}$ are also spatial-dependent.

\subsubsection{Inhomogeneous ionizing background and shrinking of islands}

The ionizing photons emitted from newly collapsed objects are gradually attenuated by small-scale absorbers, while entirely blocked when they reach a neutral island. They can travel out to a typical distance of the MFP. The ionizing background is inhomogeneous due to the inhomogeneous distribution in both sources and sinks, and the limited MFP. In the approximation that the background ionizing photons can travel freely within a distance of one MFP, the comoving number density of background ionizing photons in Eq.(\ref{Eq.ngamma_homo}) can be simplified as:
\begin{equation}\label{Eq.ngamma_inhomo}
\begin{split}
    n_{\gamma}(\boldsymbol{x}, z) \approx
\bar{n}_{\mathrm{H}}\zeta\,\left|\frac{\mathrm{d}f^\lambda_{\text{coll}}(\boldsymbol{x}, z)}{\mathrm{d}t}\right|\lambda_{\mathrm {mfp}}(\boldsymbol{x}, z)/c.
\end{split}
\end{equation}
Note that here we do not have the $\left(1-f_{\mathrm {HI}}^{\mathrm {host}}\right)$ factor as in the homogeneous model, because in the updated model the intensity of the ionizing background is only computed in ionized regions.
$\lambda_{\rm mfp}(\boldsymbol{x}, z)$ is the proper MFP, and $\lambda_{\rm mfp}(\boldsymbol{x}, z)$, $\mathrm{d}f_{\text{coll}}(\boldsymbol{x}, z)/\mathrm{d}t$, and $n_{\gamma}(\boldsymbol{x}, z)$ are now all position-dependent. Note also that in Eq.(\ref{Eq.ngamma_inhomo}) the collapse fraction should be smoothed on the scale of the local MFP, instead of the filtering scale $R$.
Assuming that ionizing background has a spectral form of $\nu^{-\eta}$, and the photoionization cross-section is $\sigma(v)=\sigma_0(\nu/\nu_0)^{-\alpha}$, in which $\sigma_0=6.3\times10^{-18}\,{\mathrm {cm}}^{2}$ and $\nu_0$ is the frequency of hydrogen ionization threshold, the
photoionization rate can be written as
\begin{equation}\label{Eq.GAMMA_HII}
\begin{split}
    \Gamma_{\rm {HII}}(\boldsymbol{x},z) = \frac{\eta\, \sigma_0}{\eta+\alpha}
    \bar{n}_{\mathrm{H}}\zeta\,\left|\frac{\mathrm{d}f^\lambda_{\text {coll}}(\boldsymbol{x},z)}{\mathrm{d}t}\right|\lambda_{\mathrm {mfp}}(\boldsymbol{x},z)(1+z)^{3}.
\end{split}
\end{equation}
We use $\eta = 5$, and $\alpha = 3$ in this paper.

The inhomogeneous recombinations and anisotropic ionizing background lead to inhomogeneous shrinking of islands. The change of the island scale $\Delta R_{\rm I}(\boldsymbol{x},z)$ is therefore also direction-dependent. Combining Eqs. (\ref{Eq.DeltaR_homo}), (\ref{Eq.ngamma_inhomo}) and (\ref{Eq.GAMMA_HII}), the scale change can be written as:
\begin{equation}\label{Eq.DeltaR_inhomo}
\begin{split}
    \Delta R_{\rm I} (\boldsymbol{x}, z) = \frac{\eta+\alpha}{4\, \eta\, \bar{n}_{\mathrm{H}}\, \sigma_{0}} \int_{z}^{z_{\mathrm{back}}} \frac{\Gamma_{\mathrm{HII}}(\boldsymbol{x}, z)}{1+\bar{n}_{\mathrm{rec}}(\boldsymbol{x}, z)}\, \frac{\mathrm{d} z}{H(z)(1+z)^{3}}.
\end{split}
\end{equation}
Following \citet{Wu2022ApJ}, we assume that an ionizing background has been set up when the reionization process enters the neutral fiber stage when the mean neutral fraction is $\bar{x}_{\rm HI}\sim 0.3$ \citep{Chen2019ApJ}, and set $z_{\mathrm {back}}$ correspondingly. The simulation switches to the two-step filtering algorithm with the island barrier when the reionization approaches the island stage at $\bar{x}_{\rm HI} < 0.2$.

\subsection{Implementation of the new islandFAST}

\begin{figure}
\centering
\includegraphics[width=1.0\textwidth]{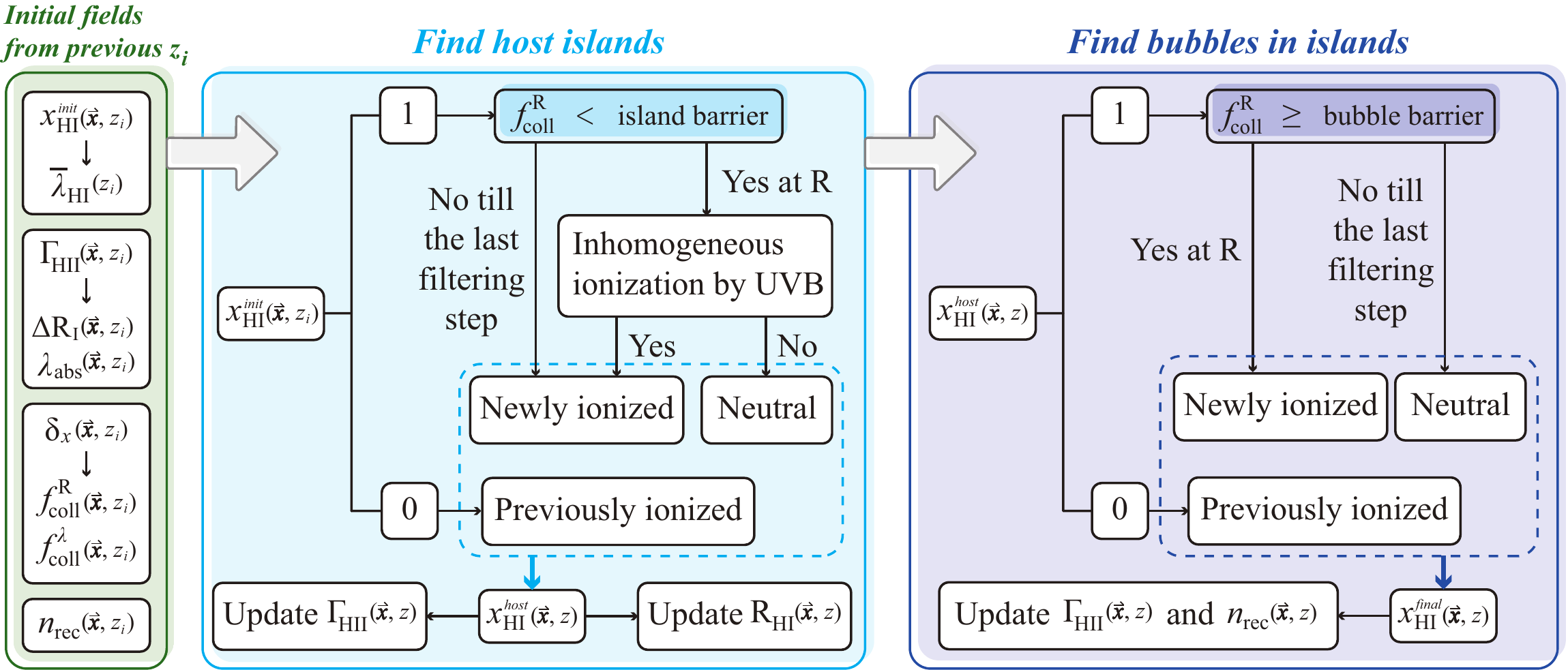}
\caption{The flow diagram of {\tt islandFAST} for each redshift. The \textit{left panel} lists the initial fields inherited from the previous redshift $z_{i}$, the \textit{middle panel} illustrates the procedure to find the host islands, and the \textit{right panel} is the procedure of finding ionized bubbles inside host islands. 
When finding the host islands, we need to consider the inhomogeneous ionizing background, so the island barrier is applied. When identifying the bubbles in islands,  no background ionizing photons would be present inside host islands, so  the bubble barrier is used.
}
\label{fig:SCHEMATIC}
\end{figure}

The basic framework of the new {\tt islandFAST} is similar to the original version, and the main steps for the island stage are illustrated in
Figure \ref{fig:SCHEMATIC}. The variables appearing in this flow diagram are summarized in Table~\ref{table:params}.

\begin{figure}
\centering
\includegraphics[width=1.0\textwidth]{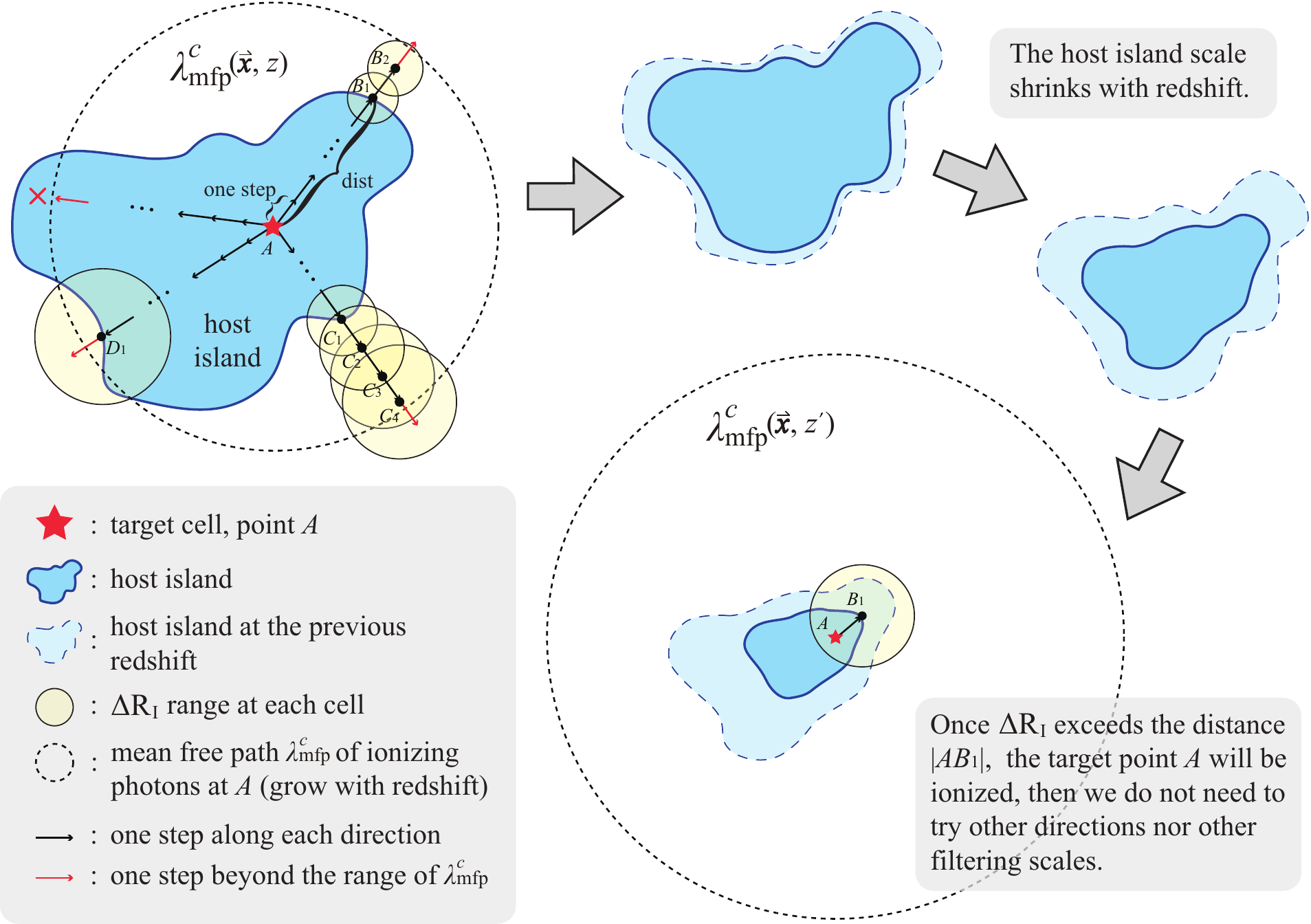}
\caption{Inhomogeneous ionization of an island by the ionizing background. To determine whether a target cell at point $A$ (represented as a red star) is ionized or not, we step outward from point $A$ in random directions, represented by little arrows in the plot. Each arrow corresponds to one step. The distance between point $A$ and the island boundary in each direction is recorded as $dist$. The blue area enclosed by the solid line is the host island, which is being ionized outside-in during every redshift interval. The light blue area with a dashed boundary is the host island at the previous redshift. The outermost dotted circle represents the comoving mean free path $\lambda^{\rm c}_{\mathrm {mfp}}$ of ionizing photons. 
Every point has its own $\Delta R_{\rm I}$ determined by the local ionizing background and the recombination numbers, which is illustrated by the size of yellow circles. After reaching the boundary of the host island (e.g. $B_{1}, C_{1}$ and $D_{1}$), we will continue going outward into the ionized region, and at each step (e.g. $C_{2}, C_{3}, C_{4}$, etc.) compare $\Delta R_{\rm I}$ with $dist$ of this direction. The target cell is identified as ionized at the current redshift once $\Delta R_{\rm I} > dist$. If the target point $A$ cannot be ionized until we step out of the MFP $\lambda_{\mathrm {mfp}}$ of the current cell (the red arrows), then this point is out of reach of the ionizing photons from the current cell, then we move to the next random direction. }
\label{fig:RANDOM}
\end{figure}

\begin{table}[h!]
\centering
\caption{Parameters  in the new islandFAST}
\label{table:params}
\begin{tabular}{cl}
\toprule
\specialrule{0em}{2pt}{2pt}
\textbf{Parameter} & \textbf{Definition} \\
\specialrule{0em}{2pt}{2pt}
\hline
\hline
\specialrule{0em}{1.5pt}{1.5pt}
$n_{\mathrm {rec}}({\boldsymbol{x}},z)$ & Recombination number in a specific cell.\\
\specialrule{0em}{1.5pt}{1.5pt}
\hline
\specialrule{0em}{1.5pt}{1.5pt}
$\delta_x({\boldsymbol{x}},z)$ & Overdensity.\\
\specialrule{0em}{1.5pt}{1.5pt}
\hline
\specialrule{0em}{1.5pt}{1.5pt}
$\Delta R_{\rm I}({\boldsymbol{x}},z)$ & Scale change of a host island that can be induced from the ionizing background at $\boldsymbol{x}$.\\
\specialrule{0em}{1.5pt}{1.5pt}
\hline
\specialrule{0em}{1.5pt}{1.5pt}
$x_{\mathrm {HI}}({\boldsymbol{x}},z)$ & \makecell[l]{Neutral fraction, whose value is either 1 (neutral) or 0 (ionized).}\\
\specialrule{0em}{1.5pt}{1.5pt}
\hline
\specialrule{0em}{1.5pt}{1.5pt}
$\Gamma_{\mathrm {HII}}({\boldsymbol{x}},z)$ & Photoionization rate of the ionizing background.\\
\specialrule{0em}{1.5pt}{1.5pt}
\hline
\specialrule{0em}{1.5pt}{1.5pt}
$\bar{\lambda}_{\mathrm {HI}}(z)$ & \makecell[l]{Mean free path limited by neutral islands, which is averaged over the whole simulation \\ box for a specific redshift.}\\
\specialrule{0em}{1.5pt}{1.5pt}
\hline
\specialrule{0em}{1.5pt}{1.5pt}
$\lambda_{\mathrm {abs}}({\boldsymbol{x}},z)$ & \makecell[l]{Mean free path limited by  small-scale absorbers.}\\
\specialrule{0em}{1.5pt}{1.5pt}
\hline
\specialrule{0em}{1.5pt}{1.5pt}
$\lambda_{\mathrm {mfp}}({\boldsymbol{x}},z)$ & \makecell[l]{The effective mean free path at a specific cell including the contribution of both islands \\ and small-scale absorbers.}\\
\specialrule{0em}{1.5pt}{1.5pt}
\hline
\specialrule{0em}{1.5pt}{1.5pt}
$f_{\mathrm {coll}}^{R}({\boldsymbol{x}},z) (f_{\mathrm {coll}}^{\lambda}({\boldsymbol{x}},z))$ & Collapse fraction that is smoothed over the filtering scale $R$ (effective mean free path $\lambda_{\mathrm {mfp}}$).\\
\specialrule{0em}{1.5pt}{1.5pt}
\hline
\specialrule{0em}{1.5pt}{1.5pt}
$dist$ & \makecell[l]{Distance from a target cell to the island boundary in a specific direction, see Figure \ref{fig:RANDOM} for \\more details.}\\
\specialrule{0em}{1.5pt}{1.5pt}
\hline
\specialrule{0em}{1.5pt}{1.5pt}
$R_{\mathrm {HI}}({\boldsymbol{x}},z)$ & \makecell[l]{Radius of host island at each cell.}\\
\specialrule{0em}{1.5pt}{1.5pt}
\hline
\end{tabular}
\end{table}

The main improvements of this work are the following:
\begin{enumerate}
    \item The MHR00 distribution is adopted for the gas density distribution of small-scale absorbers. Based on this model, inhomogeneous recombinations and the position-dependent MFP are self-consistently calculated along with the inhomogeneous ionizing background.
    \item At each filtering step, the collapse fraction is smoothed on both the filtering scale $R$ and the MFP $\lambda_{\rm mfp}$. The former is used in the barriers when identifying host islands as well as the bubbles in islands, and the latter is used for calculating the intensity of the ionizing background, as the contributing photons mainly come from sources within one MFP.
    \item Inhomogeneous ionization, or direction-dependent shrinking, of islands is incorporated when identifying the host islands. When a cell is identified as possibly within a host island, random directions are selected, and in each direction, we step outward starting from this central cell. At each step beyond the edge of the host island, one compares $\Delta R_{\rm I}({\boldsymbol{x}},z)$ of the current position with the distance between the island edge and the central cell. 
The comparison in this direction continues until an MFP is reached.
This central cell will be ionized from this direction if $\Delta R_{\rm I}({\boldsymbol{x}},z)$ is larger than the distance. One cell is considered ionized if it can be ionized by the ionizing background from any of the random directions. We have tested that 100 random directions for each neutral cell can safely result in convergent ionization fields. This procedure is illustrated in Figure~\ref{fig:RANDOM}, and is shown as the block of ``Inhomogeneous ionization by UVB'' in Figure~\ref{fig:SCHEMATIC}.
\end{enumerate}

\begin{figure}
\centering
\includegraphics[width=1.0\textwidth]{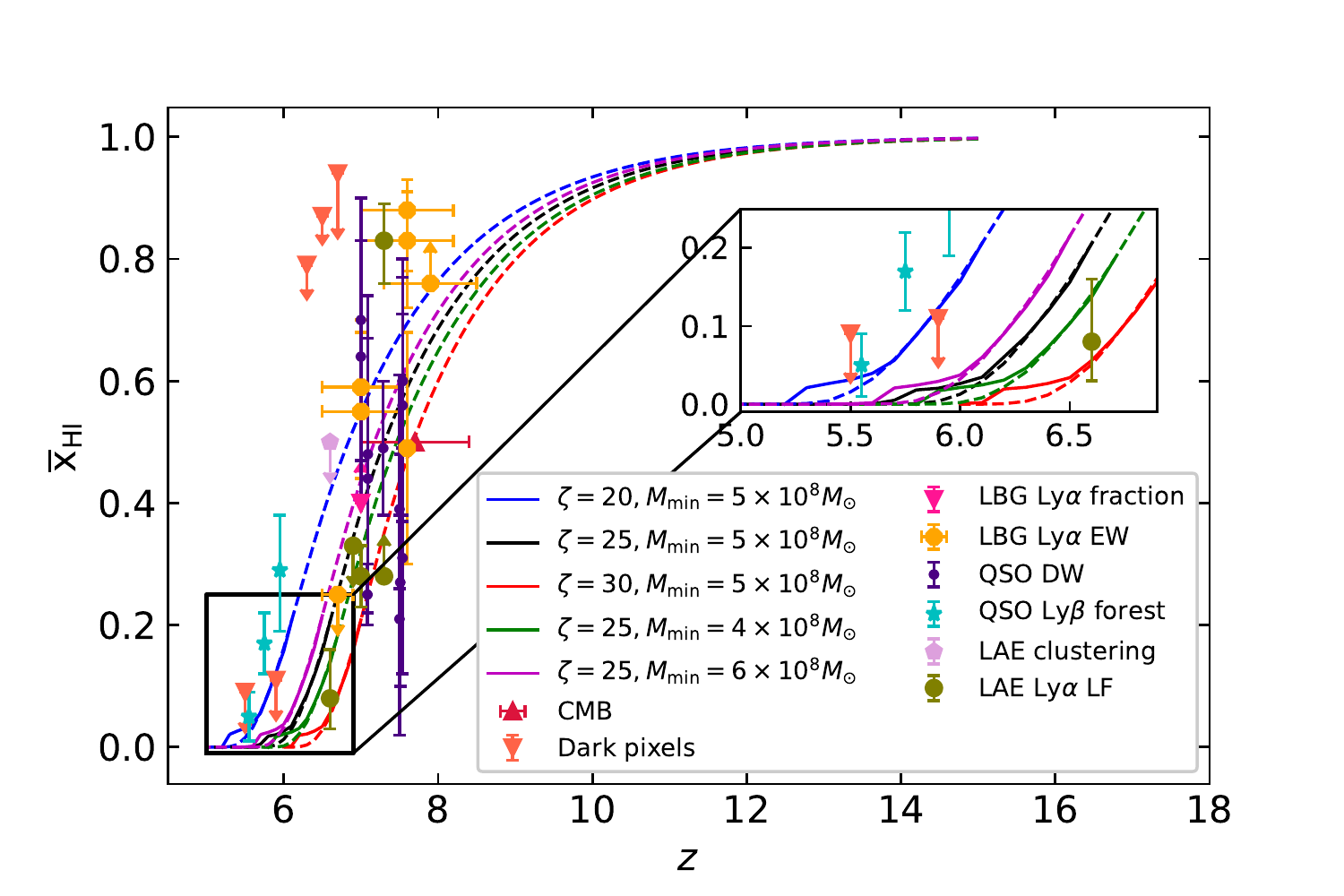}
\caption{The global reionization history of the IGM from {\tt islandFAST} (\textit{solid lines}) and {\tt 21cmFAST} (\textit{dashed lines}) semi-numerical simulations, for various combinations of $\zeta$ and $M_{\mathrm {min}}$. The black lines are for the fiducial parameters of $\zeta = 25$ and $M_{\mathrm {min}} = 5\times 10^{8}M_{\odot}$. 
Various observational constraints are also plotted for comparison: \textit{Red-} the CMB optical depth measurement from \citet{Planck2020}. \textit{Orange-} dark pixels on quasar spectra from \citet{McGreer2015MNRAS} and \citet{Jin2023ApJ}. \textit{Magenta-} the Ly$\alpha$ fraction of LBGs  at $z = 7$ \citep{Mesinger2015MNRAS}. \textit{Yellow-} the Ly$\alpha$ EW distribution at $z = 7$ \citep{Mason2018ApJ,Whitler2020MNRAS}, $z = 7.6$ \citep{Hoag2019ApJ,Jung2020ApJ}, $z = 6.7$ and 7.6 \citep{Bolan2022MNRAS}, and $z = 7.9$ \citep{Mason2019MNRAS}. \textit{Purple-} quasar damping wings from \citet{Banados2018Natur, Greig2019MNRAS, Greig2022MNRAS, Davies2018ApJ, Durovckov2020MNRAS, Wang2020ApJ} and \citet{Yang2020ApJ}. \textit{Cyan-} the Ly$\beta$ forest in quasar spectra by \citet{Zhu2022ApJ}. \textit{Light pink-} LAE clustering at $z = 6.6$ \citep{Sobacchi2015MNRAS}. \textit{Green-} the Ly$\alpha$ LFs by \citet{Morales2021ApJ}, \citet{Goto2021ApJ}, and \citet{Wold2022ApJ}.}
\label{fig:XH_CON}
\end{figure}

We run both {\tt 21cmFAST} and the new {\tt islandFAST} with the same box size of 500 comoving $\Mpc$. The initial density field has a resolution of $1500^{3}$, and the evolved density and ionization fields are smoothed to have $N = 500^{3}$ cells. 
We vary $\zeta$ and the minimum halo mass for contributing ionizing photons $M_{\rm min}$, and choose a number of combinations for which the reionization history is consistent with existing observations, including the Thomson optical depth measurement to the CMB \citep{Planck2020}, the dark pixels statistics on quasar spectra \citep{McGreer2015MNRAS, Jin2023ApJ}, the fraction of Lyman-break galaxies (LBGs) with a strong Ly$\alpha$ line (the Ly$\alpha$ fraction, \citealt{Mesinger2015MNRAS}), the Ly$\alpha$ equivalent-width (EW) distribution of LBGs \citep{Mason2018ApJ, Mason2019MNRAS,Hoag2019ApJ,Jung2020ApJ,Whitler2020MNRAS,Bolan2022MNRAS}, the Ly$\alpha$ damping wings of quasars \citep{Banados2018Natur, Greig2019MNRAS, Greig2022MNRAS, Davies2018ApJ, Durovckov2020MNRAS, Wang2020ApJ, Yang2020ApJ}, the dark gaps in Ly$\beta$ forest \citep{Zhu2022ApJ}, the clustering of Ly$\alpha$ emitters (LAEs) \citep{Sobacchi2015MNRAS} as well as the Ly$\alpha$ luminosity functions (LFs) \citep{Morales2021ApJ, Goto2021ApJ, Wold2022ApJ}. 
The ionization histories for several combinations of \{$\zeta , M_{\mathrm {min}}$\} are shown in
Figure \ref{fig:XH_CON}, with solid lines for {\tt islandFAST} and dashed lines for {\tt 21cmFAST} respectively. 
The Thomson optical depth to the CMB for these parameter combinations is listed in Table \ref{table:TAUE}, all within the 1-$\sigma$ constraint from the Planck observation \citep{Planck2020}.
In this work, we take \{$\zeta = 25$, $M_{\mathrm {min}} = 5\times 10^{8}M_{\odot}$\} as the fiducial model, which corresponds to $\tau = 0.055$. 

\begin{table}[h!]
\centering
\caption{The Thomson optical depth to the CMB in two models for three values of $\zeta$, assuming $M_{\rm min} = 5\times 10^8 M_\odot$.}
\label{table:TAUE}
\begin{tabular}{ccccc}
\hline
\specialrule{0em}{2pt}{2pt}
\multicolumn{2}{c}{\textbf{Model}} & \bm{$\zeta = 20$} & \bm{$\zeta = 25$} & \bm{$\zeta = 30$}\\
\specialrule{0em}{2pt}{2pt}
\hline
\hline
\specialrule{0em}{1.5pt}{1.5pt}
\multirow{2}{*}{islandFAST}& inhomogeneous& 0.050& 0.055& 0.058\\ \cline{2-2}
& homogeneous & 0.051& 0.056& 0.059 \\
\specialrule{0em}{1.5pt}{1.5pt}
\hline
\specialrule{0em}{1.5pt}{1.5pt}
\multicolumn{2}{l}{21cmFAST}& 0.050& 0.055& 0.058\\
\specialrule{0em}{1.5pt}{1.5pt}
\hline
\specialrule{0em}{1.5pt}{1.5pt}
\multicolumn{2}{l}{Planck 2020}& \multicolumn{3}{c}{0.054 $\pm$ 0.007}\\
\specialrule{0em}{1.5pt}{1.5pt}
\hline
\end{tabular}
\end{table}

As expected, a higher ionizing efficiency $\zeta$ results in faster reionization, and a smaller collapse threshold $M_{\mathrm {min}}$ leads to earlier beginning of reionization.
However, the reionization process is more delayed at the late EoR in {\tt islandFAST}, as compared to the process simulated by {\tt 21cmFAST}. 
Besides the different topologies assumed by the two models and the different filtering algorithms, this is partly because 
the {\tt islandFAST} distinguishes $f_{\rm coll}^{R}$ in the barriers and $f_{\rm coll}^{\lambda}$ in the formula for the ionizing background, and only the collapsed objects within a distance of $\lambda_{\mathrm {mfp}}$ are counted when computing the number density of background ionizing photons.

\section{Effects of inhomogeneous recombinations}
\label{sect:nrec}

\begin{figure}
\centering
\includegraphics[width=1.0\textwidth]{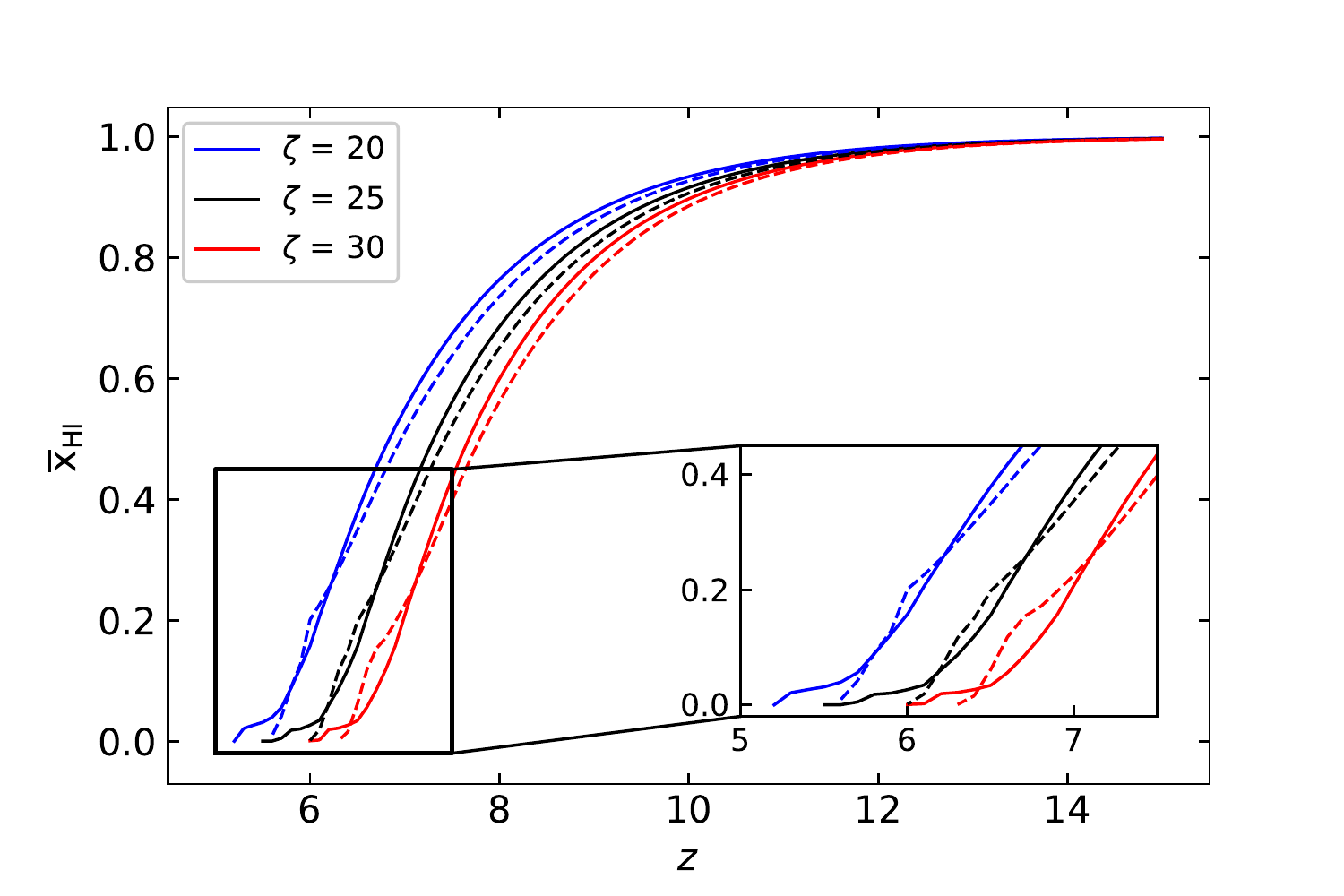}
\caption{The global reionization history of the IGM  $\bar{x}_{\mathrm {HI}}(z)$ that derived from the inhomogeneous recombination model (\textit{solid lines}) and the spatially homogeneous recombination model (\textit{dashed lines}). The blue, black and red lines correspond to $\zeta = 20, 25,$ and $30$ respectively. 
}
\label{fig:XH_EVO_INHO}
\end{figure}

\begin{figure}
\centering
\includegraphics[width=1.0\textwidth]{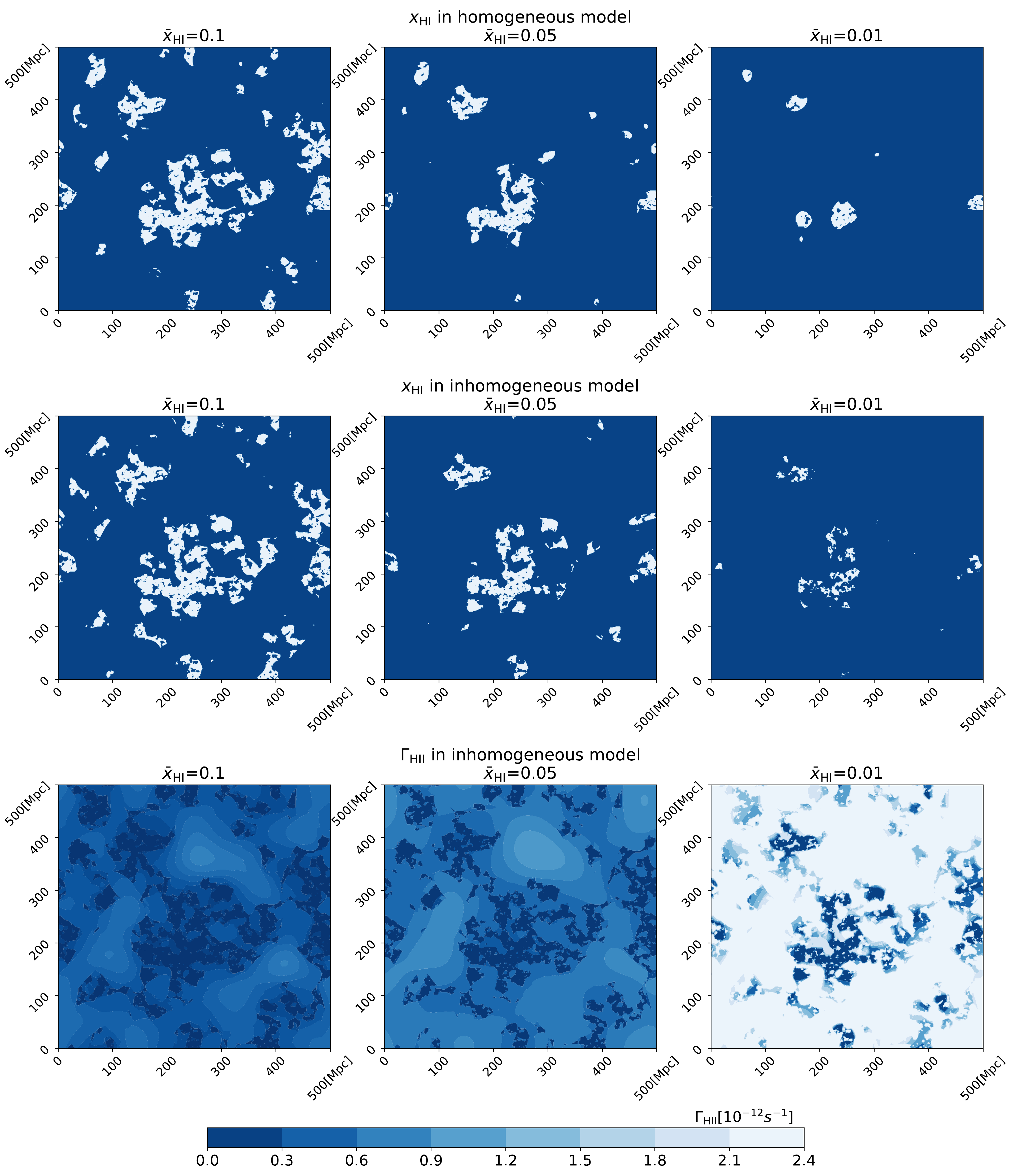}
\caption{Slices of ionization fields from {\tt islandFAST} with the homogeneous model (\textit{upper three panels}) and the inhomogeneous model (\textit{middle three panels}), and slices of photoionization rate of the ionizing background from the inhomogeneous model (\textit{bottom three panels}). Fiducial parameters of $\zeta = 25$ and $M_{\mathrm{min}} = 5\times 10^{8} M_{\odot}$ are adopted. The left, central, and right columns correspond to $\bar{x}_{\mathrm {HI}} = 0.10, 0.05$, and $0.01$, respectively. In the ionization fields, the ionized ($x_{\mathrm {HI}} = 0$) and neutral ($x_{\mathrm {HI}} = 1$) regions are represented by dark blue and light blue respectively. The photoionization rate is shown with the color bar at the bottom.}
\label{fig:SLICE_INHO}
\end{figure}

\begin{figure}
\centering
\includegraphics[width=1.0\textwidth]{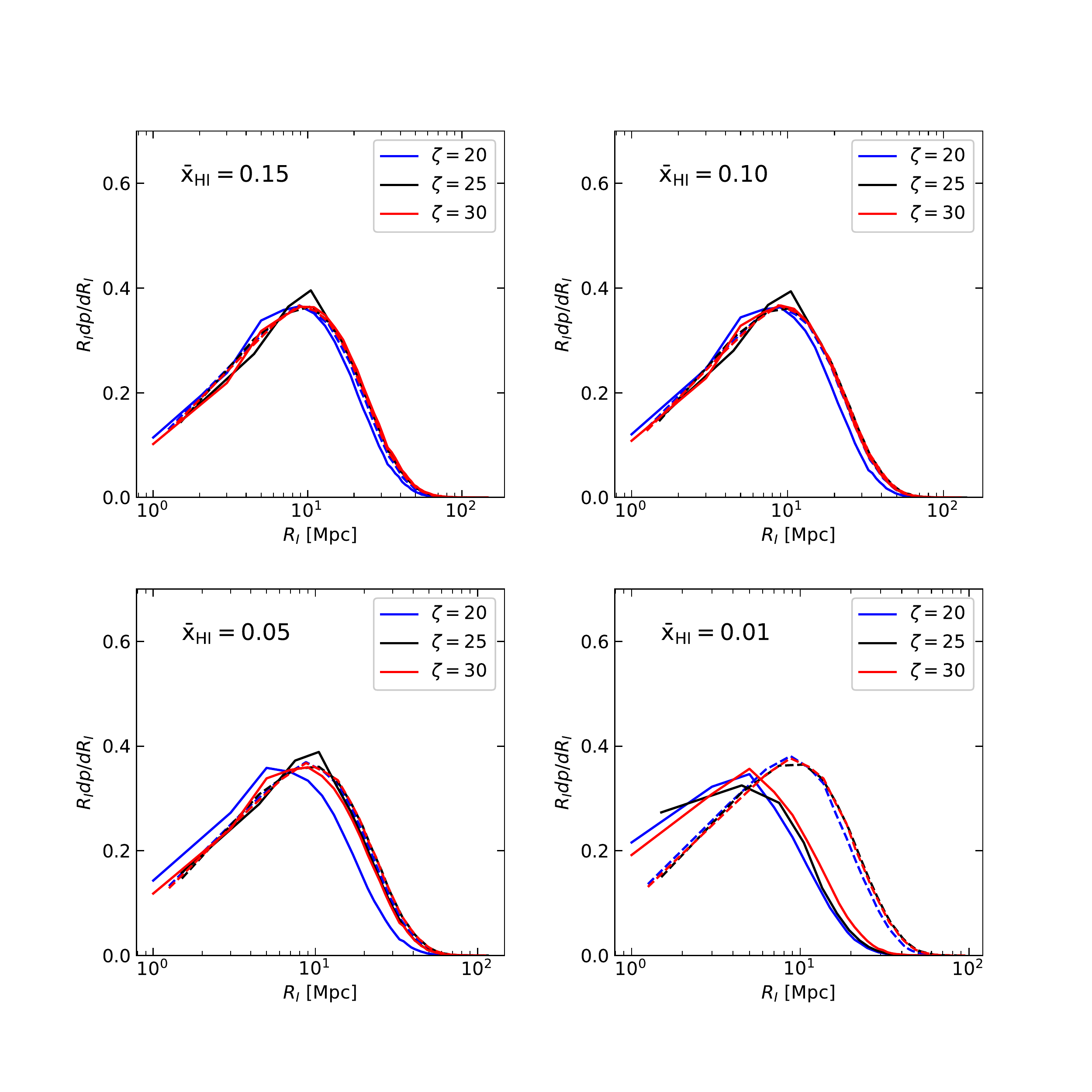}
\caption{Size distribution of neutral islands during the late stages of reionization at four mean neutral fractions. The \textit{solid} and \textit{dashed} lines are the results predicted from the inhomogeneous model and the homogeneous model respectively. We assume $M_{\mathrm{min}} = 5\times 10^{8} M_{\odot}$, and in each panel, the blue, black, and red lines are for $\zeta = 20$, $\zeta = 25$, and $\zeta = 30$, respectively. }
\label{fig:SIZE_INHO}
\end{figure}

\begin{figure}
\centering
\includegraphics[width=1.0\textwidth]{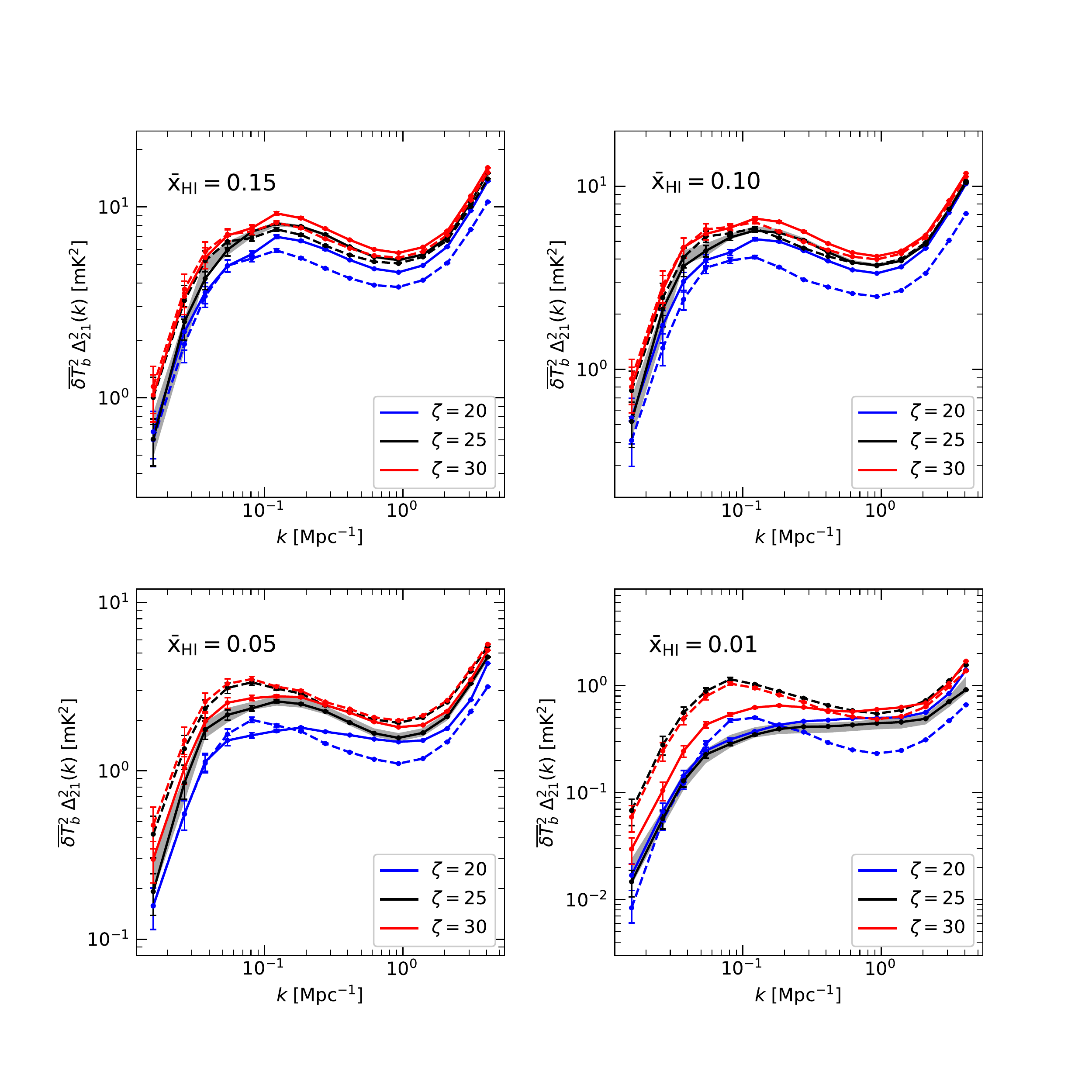}
\caption{21 cm power spectra with statistical errors at four stages of late EoR, with mean neutral fractions of $\bar{x}_{\mathrm {HI}} = 0.15, 0.10, 0.05$ and $0.01$. The blue, black, and red lines show the results from the inhomogeneous model (\textit{solid line}) and the homogeneous model (\textit{dashed line}) with $\zeta =$ 20, 25, and 30, respectively, and $M_{\mathrm{min}} = 5\times 10^{8} M_{\odot}$ is adopted.
In each plot, the grey shade shows the root-mean-square deviation from 10 realizations of the fiducial model.}
\label{fig:21CM_INHO}
\end{figure}

With the improved algorithm, we investigate the effects of inhomogeneous recombinations and an inhomogeneous ionizing background. The new {\tt islandFAST} is called the ``inhomogeneous model'' hereafter. For comparison, we also run another simulation called the ``homogeneous model'' with the same filtering process and same parameters, but a uniform recombination number that is equal to the averaged value from the inhomogeneous model is adopted for each redshift.

Assuming $M_{\mathrm {min}} = 5\times 10^{8}M_{\odot}$, the ionization histories for three different ionizing efficiencies of $\zeta =$ 20, 25 and 30 are illustrated in Figure \ref{fig:XH_EVO_INHO} by blue, black and red lines, respectively. 
The solid lines correspond to the results from the inhomogeneous model, while the dashed lines are for the homogeneous model.  
Generally, with the inhomogeneous recombinations, the overall reionization history is only slightly modified. During the early stages, when a  large fraction of the IGM is neutral, the averaged recombination number under-estimates the recombination number in ionized regions, and the reionization process is slightly faster in the homogeneous model. 
When the reionization progressed to the neutral fiber stage (for the reionization stages, c.f. \citealt{Chen2019ApJ}) and the island stage at $\bar{x}_{\mathrm {HI}} \lesssim 0.3$, the ionization process in the inhomogeneous model becomes faster. This is because the neutral fibers and islands are generally located in underdense regions where the recombination number is lower than the average, and hence the shrinking process of neutral regions would be faster than in the homogeneous model. 
At the very last stage of reionization, only small islands far from ionizing sources are left over, then the ionization process is delayed again in the inhomogeneous model in which the ionizing background intensity is much lower near those islands.

Slices of the ionization field during the island stage of reionization are shown in Figure \ref{fig:SLICE_INHO}, with the upper panels from the homogeneous model, and the middle panels from the inhomogeneous model. The lower panels are the 
corresponding $\Gamma_{\rm HII}$ slices from the inhomogeneous model. 
The three columns are for three epochs corresponding to mean neutral fractions of $\bar{x}_{\mathrm {HI}} =$ 0.10, 0.05, and 0.01, from left to right respectively.
In the ionization fields, ionized regions are shown by dark blue and neutral islands are in light blue. 
By assuming that the ionizing photons are dominated by UV photons, the ionized fraction is either 0 or 1 for each pixel, but we take into account the ``bubbles-in-island'' effect to account for partially-ionized islands.
From the ionization slices, it is found that with the inhomogeneous recombinations, the neutral islands fragment into more small pieces. At the same mean neutral fraction of the universe, the typical scale of islands in the inhomogeneous model would be smaller than the one in the homogeneous model. 
However, the difference in the morphology of ionization field only becomes obvious at the very late stage of reionization.
The ionizing background intensity field roughly follows the ionization field, regulated by the density field.
At the end of reionization, the photoionization rate near neutral islands is typically small, resulting in slower ionization of the islands left over.

The morphological difference in the ionization field can be characterized by the size distribution of neutral islands. We apply the mean-free-path algorithm \citep{Mesinger2007ApJ} to extract the island size distribution for three values of ionizing efficiency parameter $\zeta$ as shown in Figure~\ref{fig:SIZE_INHO}, assuming $M_{\rm min} = 5\times 10^8 M_{\odot}$. The solid lines are for the inhomogeneous model, while the dashed lines are for the homogeneous model.
The inhomogeneous model makes little difference in the size distribution of islands down to a mean neutral fraction of $\bar{x}_{\mathrm {HI}} \sim 0.1$. The typical scale of neutral islands is around 10 comoving Mpc at $\bar{x}_{\mathrm {HI}} \gtrsim 0.1$. When the mean neutral fraction drops to $\bar{x}_{\mathrm {HI}} \sim 0.05$, the neutral regions tend to fragment into smaller islands in the inhomogeneous model, while the homogeneous model would overpredict the typical size of islands.  When the reionization is about to be completed, the typical scale of islands in the inhomogeneous model is about 5 Mpc, significantly smaller than the prediction from the homogeneous model. 
Interestingly, by comparing results for a larger range of astrophysical parameters, it is found that the size distribution of neutral islands, when compared at the same $\bar{x}_{\mathrm {HI}}$, is not sensitive to the source parameters for $15<\zeta<35$ and $10^8 M_{\odot} < M_{\rm min} < 10^9 M_{\odot}$. Therefore, the size distribution of neutral islands at the end of reionization can be potentially a diagnostics for inhomogeneous recombinations and an inhomogeneous ionizing background.

We also investigate the effect of inhomogeneous recombinations on the 21 cm power spectrum. The differential 21 cm brightness temperature of neutral hydrogen against the CMB can be calculated  by (e.g. \citealt{Furlanetto2006PhR}):
\begin{equation}
\begin{split}
    \begin{aligned}
    \delta T_{\mathrm{b}}(\nu)&\approx 27 x_{\mathrm{H} \mathrm{I}}\left(1+\delta_{\mathrm{nl}}\right)\left(\frac{H}{\mathrm{~d}
    v_{\mathrm{r}} / \mathrm{d} r+H}\right)\left(1-\frac{T_\gamma}{T_{\mathrm{S}}}\right)
    \,\left(\frac{1+z}{10} \frac{0.15}{\Omega_{\mathrm{m}} h^2}\right)^{1 / 2}\left(\frac{\Omega_{\mathrm{b}} h^2}{0.023}\right) \mathrm{mK},
    \end{aligned}
\end{split}
\end{equation}
where $x_{\rm HI}$ is the neutral fraction of hydrogen gas, $\delta_{\mathrm {nl}}$ is the evolved density contrast, $H(z)$ is the Hubble parameter, $\mathrm{d}v_{\mathrm {r}}/\mathrm{d}r$ is the velocity gradient projected to the line of sight in comoving coordinates, and $T_{S}$ and $T_{\gamma}$ are the spin temperature and the CMB brightness temperature respectively.
We assume $T_{S} \gg T_{\gamma}$ in this work as the gas is probably heated during the late EoR \citep{Pritchard2007MNRAS, Chen2008ApJ}. 

The 21 cm power spectra with statistical errors as expected from the inhomogeneous model and the homogeneous model are shown in Figure \ref{fig:21CM_INHO} with solid and dashed lines respectively. 
The four panels show the 21 cm power spectra at $\bar{x}_{\rm HI} = 0.15, 0.10, 0.05$, and 0.01, respectively.
In each set of lines, different colors are for different ionizing efficiency parameters.
For the fiducial model with $\zeta = 25$, the grey shade corresponds to the root-mean-square deviation from 10 random realizations of the simulation.
When the universe enters the island stage of reionization, as seen from the upper-left plot for $\bar{x}_{\mathrm {HI}} = 0.15$, the inhomogeneous model predicts higher 21 cm power spectrum on small scales, as compared to the homogeneous model. 
As $\bar{x}_{\mathrm {HI}}$ decreases, the homogeneous model generally predicts higher power on large scales, as a result of the lower number of larger islands.
However, unlike the island size distribution, the ionizing efficiency $\zeta$ significantly affects the amplitude of 21 cm power spectrum,
so it is crucial to account for the inhomogeneous recombinations in order to precisely infer astrophysical parameters.  
Generally, at the end of reionization, the inhomogeneous model predicts flatter 21 cm power spectra than the homogeneous model.

\section{Comparison of Models}
\label{sect:model}

\begin{figure}
\centering
\includegraphics[width=1.0\textwidth]{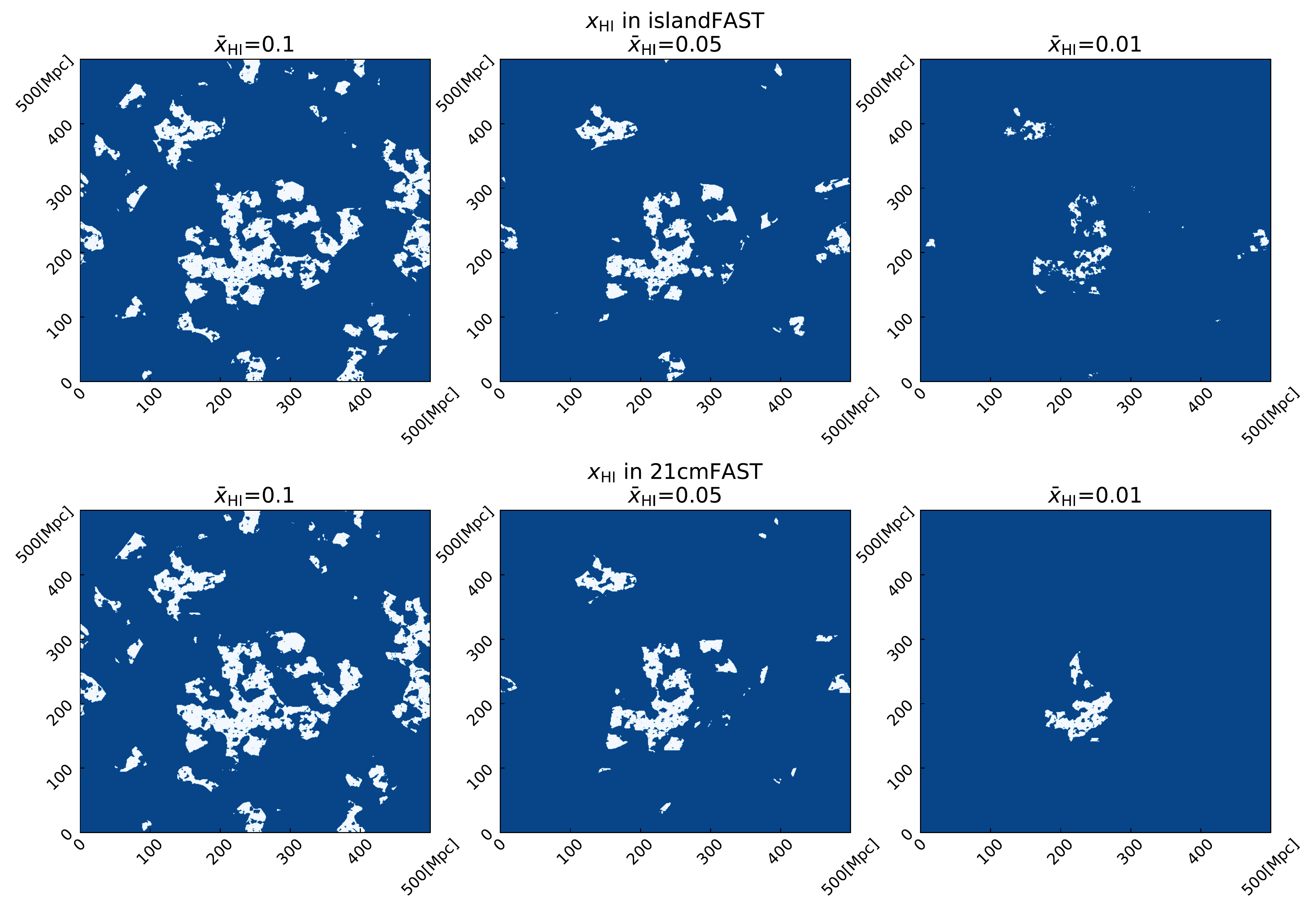}
\caption{Slices of ionization fields from {\tt islandFAST} (\textit{upper three panels}) and those from {\tt 21cmFAST} (\textit{lower three panels}) assuming $\zeta = 25$, and $M_{\mathrm {min}} = 5\times 10^{8}M_{\odot}$. 
The three columns from left to right correspond to $\bar{x}_{\mathrm {HI}} = 0.10, 0.05$ and $0.01$, respectively.}
\label{fig:SLICE_21IS}
\end{figure}

\begin{figure}
\centering
\includegraphics[width=1.0\textwidth]{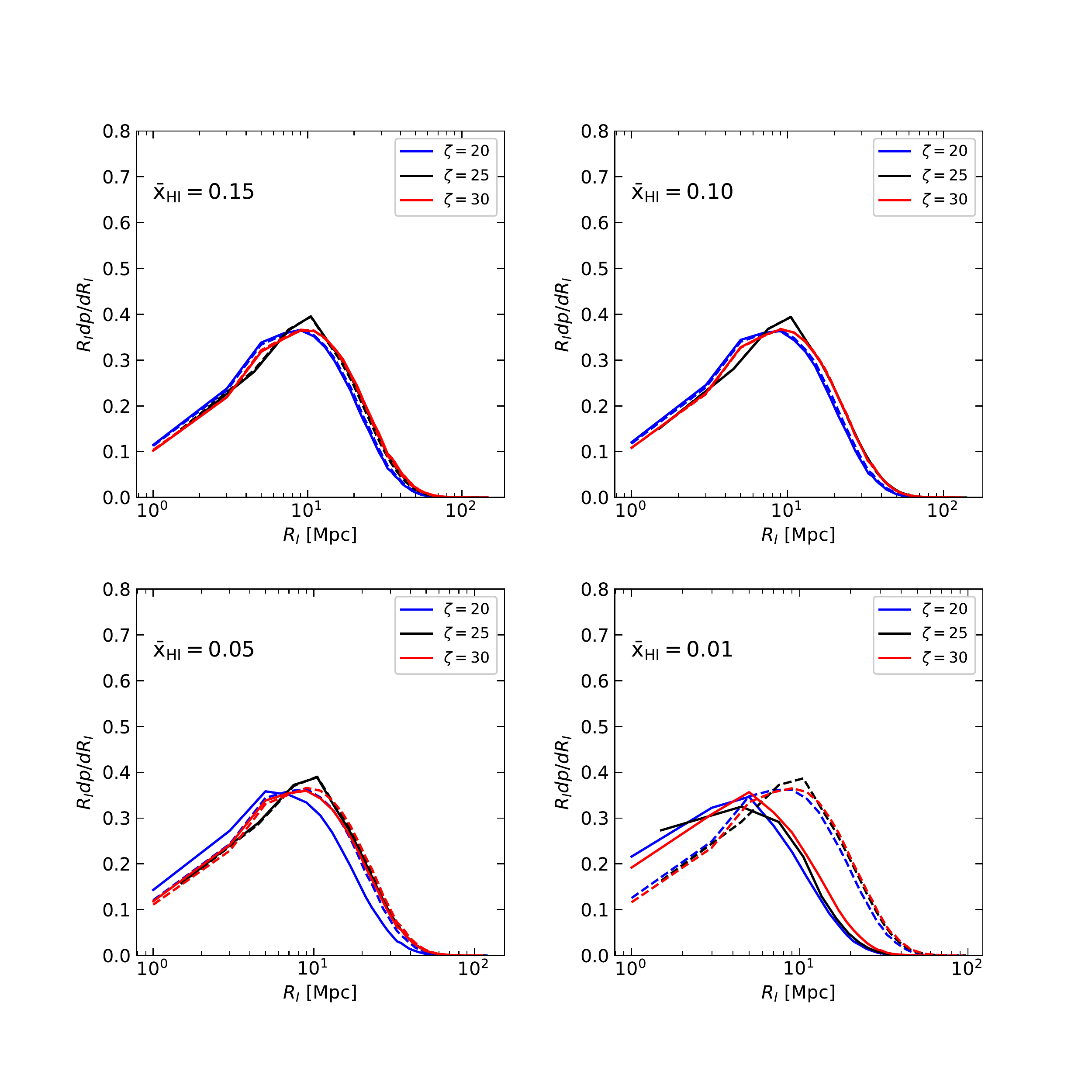}
\caption{Size distribution of neutral islands from {\tt islandFAST} (\textit{solid lines}) and {\tt 21cmFAST} (\textit{dashed lines}) at four mean neutral fractions ($\bar{x}_{\mathrm {HI}} = 0.15, 0.10, 0.05$ and $0.01$). 
The blue, black, and red lines are for $\zeta = 20, 25, 30$ respectively. Here we use the fiducial value $M_{\mathrm {min}} = 5\times 10^{8}M_{\odot}\}$.
}
\label{fig:SIZE_21IS}
\end{figure}

\begin{figure}
\centering
\includegraphics[width=1.0\textwidth]{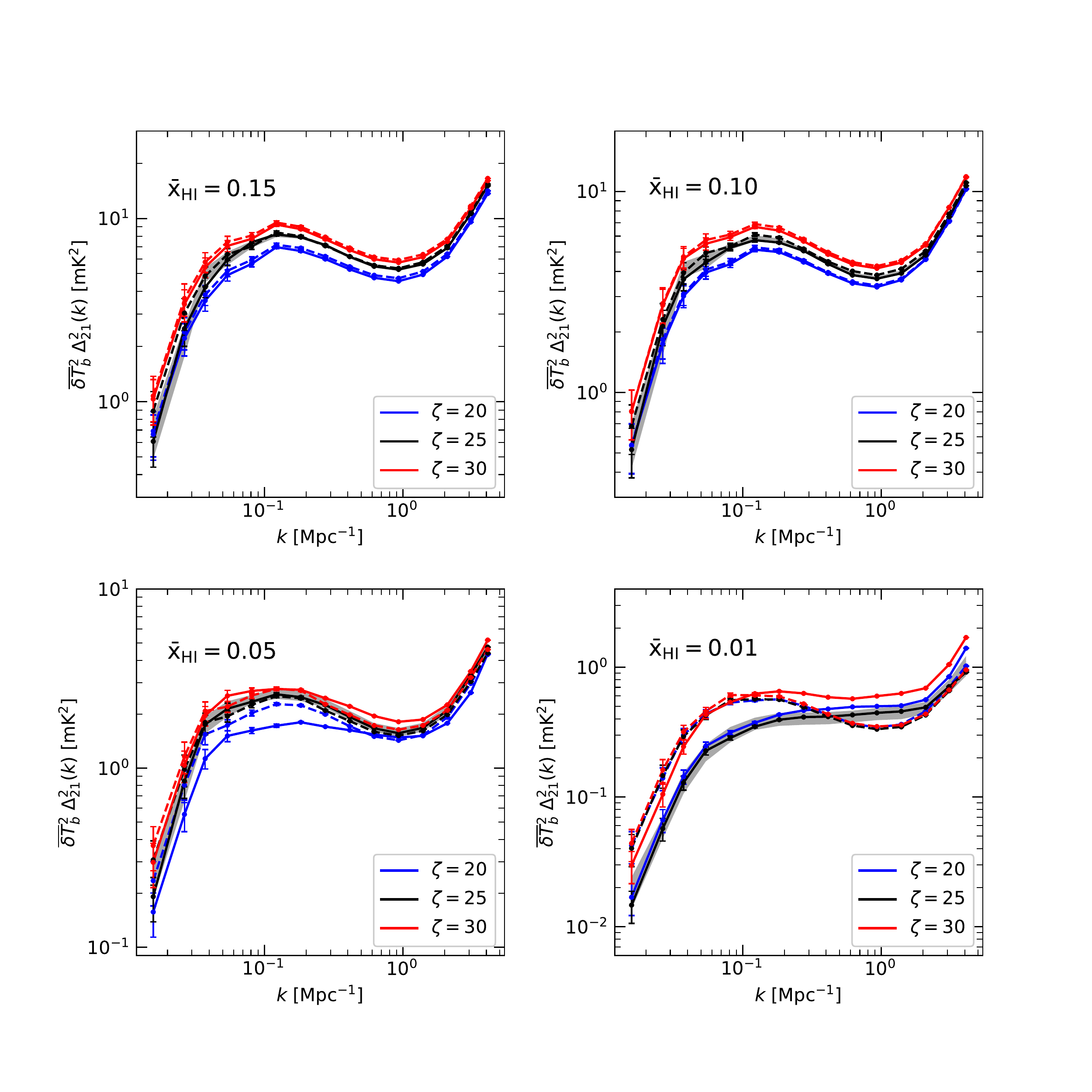}
\caption{The 21 cm power spectra with statistical errors from {\tt islandFAST} (\textit{solid lines}) and {\tt 21cmFAST} (\textit{dashed lines}) at four mean neutral fractions ($\bar{x}_{\mathrm {HI}} = 0.15, 0.10, 0.05$ and $0.01$) for the same parameters in the last figure. 
In each plot, the grey shade shows the root-mean-square deviation from 10 realizations of the fiducial model.
}
\label{fig:21CM_21IS}
\end{figure}

From the above analysis, it is seen that incorporating inhomogeneous recombinations and an inhomogeneous ionizing background in the model can result in morphological change in the ionization field during the island stage. On the other hand, the basic assumption on the ionization topology and the detailed algorithm in the modeling also affect the ionization field morphology, even though the global ionization history could be similar.
The {\tt islandFAST} improves the performance of the excursion set theory for the late EoR by adopting the topology of isolated islands, which is more appropriate for the island stage of reionization. The corresponding filtering algorithm for the ionization field incorporates the direct interaction between the inhomogeneous ionizing background and the anisotropic shrinking of the neutral islands. The effect of an inhomogeneous ionizing background is not limited to a modulation on the spatial-dependent recombination rate as in {\tt 21cmFAST}. In addition, by calculating also the collapse fraction smoothed on scales of MFP, the computation of ionizing background in {\tt islandFAST} 
is decoupled from the filtering scale $R$, and it is more physical to count only the ionizing photons within a distance of
$\lambda_{\rm mfp}$. With these improvements, there are apparent changes in the resulting ionization field morphology at the end of EoR, compared with the results from either the earlier version of {\tt islandFAST} (homogeneous model) or the  
{\tt 21cmFAST}.

Here we present the difference
in the ionization field by comparing the results from the new {\tt islandFAST} with those from {\tt 21cmFAST}. Both simulations have the same model parameters for the ionizing sources, the distribution of small-scale absorbers (recombinations), and the inhomogeneous ionizing background, and the ionization histories are all consistent with observations.
Figure \ref{fig:SLICE_21IS} shows slices of the ionization field from the two models at three different mean neutral fractions, with the upper panels for {\tt islandFAST} and the lower panels for {\tt 21cmFAST} respectively. The fiducial parameter set of \{$\zeta=25, M_{\mathrm {min}} = 5\times 10^{8}M_{\odot}$\} is adopted. It is seen that the morphology of the ionization fields are very similar between the two models until the epoch approaching the end of reionization. At $\bar{x}_{\rm HI} \sim 0.01$, the {\tt 21cmFAST} predicts a lower number of relatively larger islands as compared to the {\tt islandFAST}, although both models include inhomogeneous  recombinations and ionizing background.

The size distributions of neutral islands at four different ionization fractions during the late EoR are shown in Figure \ref{fig:SIZE_21IS}. Note that this is a relative size distribution plot, and the total number density and volume fraction of the neutral islands decrease with the decreasing total ionization fraction which are not shown in this plot. The dashed lines are the results from {\tt 21cmFAST} and the solid lines are from {\tt islandFAST},
different colors correspond to different parameter values as indicated in the legends. 

As have been noted in earlier papers (c.f. \citealt{Xu2017ApJ,Wu2022ApJ}), there is little evolution in the island size distribution for the {\tt 21cmFAST} result, while in {\tt islandFAST}, the island size distribution starts to evolve to smaller scales at $\bar{x}_{\rm HI}\lesssim 0.05$
if there are numerous small-scale absorbers.
For the parameters adopted here, which are consistent with existing observations,
the typical size of islands is of the order of 10 comoving Mpc for {\tt 21cmFAST} throughout the late EoR.  For {\tt islandFAST}, the island size is also about 10 comoving Mpc when $\bar{x}_{\rm HI}\gtrsim 10\%$. When $\bar{x}_{\mathrm {HI}} \sim 0.01$, however, the neutral islands fragment into much smaller ones in the island model, and the typical island size decreases to $\sim 5$ comoving Mpc. This agrees with our intuition that the ionizing background and the ``bubbles-in-island'' effect help to break the 
last-remaining neutral islands in the lowest density regions, where there is little star formation, into smaller ones. This is a significantly different prediction of the {\tt islandFAST} model. When compared at the same neutral fraction, the island size distribution is not sensitive to the parameter of $\zeta$. We have also checked that the results are not sensitive to the minimal collapse mass $M_{\rm min}$, though this is not shown here.

The difference in morphology is also reflected by the 21 cm power spectrum, as shown in Figure \ref{fig:21CM_21IS} for the same model parameters. The solid lines are from {\tt islandFAST} and the dashed lines are from {\tt 21cmFAST}. As expected, the two models predict similar 21 cm power spectra at $\bar{x}_{\rm HI} \gtrsim 0.10$, and the deviation starts to appear at $\bar{x}_{\mathrm {HI}} \sim 0.05$. At the very end of reionization, i.e. at $\bar{x}_{\rm HI} \sim 0.01$, the {\tt 21cmFAST} predicts higher 21 cm power spectra on large scales, while the {\tt islandFAST} predicts higher power on small scales. This is consistent with the different features as seen in the visual morphology and the size distributions in Figure~\ref{fig:SLICE_21IS} and Figure~\ref{fig:SIZE_21IS} respectively. Due to the different filtering algorithms and the ``bubbles-in-island'' effect, the islands tend to fragment into more small pieces at the end of reionization in the island model. These small islands contribute significantly to the 21 cm power spectrum on small scales. Note that for the same model, the astrophysical parameter of $\zeta$ (and $M_{\rm min}$) also affects the amplitude of 21 cm power spectrum, so this model bias will affect the correct extraction of the astrophysical parameters such as $\zeta$ and $M_{\rm min}$ if it is not properly accounted for. 
Also, given the sparsity of neutral islands approaching the end of reionization, a sufficiently large simulation box with enough resolution would be needed when extracting the astrophysical parameters.

\section{Summary}
\label{sect:sum}

In this work, we improved the semi-numerical simulation {\tt islandFAST} for the hydrogen reionization.
By introducing a sub-grid model for the small-scale density distribution, we incorporated inhomogeneous recombinations and a consistent model for an inhomogeneous ionizing background. 
In addition, we modeled the inhomogeneous ionization process of islands, and developed the direction-dependent shrinking algorithm for islands' evolution.
The new version of {\tt islandFAST} can predict the reionization history consistent with a variety of observational constraints.
For the same parameters and the same prescription of inhomogeneous recombinations, {\tt 21cmFAST} predicts a larger typical size of islands at $\bar{x}_{\rm HI}\lesssim 0.05$ as compared to the improved {\tt islandFAST}.

Using the updated {\tt islandFAST}, we have studied the effects of the inhomogeneity in both the recombinations and the ionizing background. 
We found that in the inhomogeneous model the completion of reionization is delayed compared to the homogeneous model,
mainly because the intensity of ionizing background is lower near the under-dense islands.
Such locally-delayed  ionization scenario is consistent with the patchyness of the reionization process \citep{Becker2015MNRAS} and the presence of long dark gaps down to $z \simeq 5.3$ \citep{ZhuYongda2021}.
Moreover, the typical island scale is smaller in the inhomogeneous model, especially at the end of reionization. 
This can be reflected by the island size distribution, as well as in the 21 cm power spectrum.
The island size distribution is a better diagnostic statistics, as the 21 cm power spectrum is more sensitive to astrophysical parameters.

\normalem
\begin{acknowledgements}
This work has been supported by National Key R\&D Program of China No. 2018YFE0120800, the National Natural Science Foundation of China grant No. 11973047, National Key R\&D Program of China No. 2022YFF0504300, and National SKA Program of China Nos. 2020SKA0110401, 2020SKA0110402.

\end{acknowledgements}

\bibliographystyle{raa}

\bibliography{island}

\end{document}